# Spin-split band hybridization in graphene proximitized with α-RuCl₃ nanosheets

*Soudabeh Mashhadi[†◊], Youngwook Kim[†‡◊], Jeongwoo Kim[§∥], Daniel Weber[⊥], Takashi Taniguchi[#], Kenji Watanabe[#], Noejung Park[∥], Bettina Lotsch[†], Jurgen H. Smet[†], Marko Burghard[†], and Klaus Kern[†∇]*

[†]Max-Planck-Institut für Festkörperforschung, Heisenbergstrasse 1, D-70569 Stuttgart, Germany

[‡]Department of Emerging Materials Science, DGIST, 333 Techno-Jungang-daero, Hyeonpung-Myun, Dalseong-Gun, Daegu, 42988 Korea

[§]Department of Physics, Incheon National University, Incheon 22012, Republic of Korea

[∥]Department of Physics, Ulsan National Institute of Science and Technology, UNIST-gil 50, Ulsan, 44919, Republic of Korea

[⊥]Department of Chemistry and Biochemistry, The Ohio State University, Columbus, OH 43210, USA

[#]National Institute for Materials Science, 1-1 Namiki, Tsukuba, 305-0044, Japan

[∇]Institut de Physique, Ecole Polytechnique Fédérale de Lausanne, CH-1015 Lausanne, Switzerland

[◊]These authors contributed equally

* email: Sh.Mashhadi@fkf.mpg.de

**Abstract:**

**Proximity effects induced in the 2D Dirac material graphene potentially open access to novel and intriguing physical phenomena. Thus far, the coupling between graphene and ferromagnetic insulators has been experimentally established. However, only very little is known about graphene's interaction with antiferromagnetic insulators. Here, we report a low temperature study of the electronic properties of high quality van der Waals heterostructures composed of a single graphene layer proximitized with $\alpha$-RuCl$_3$. The latter is known to become antiferromagnetically ordered below 10 K. Shubnikov de Haas oscillations in the longitudinal resistance together with Hall resistance measurements provide clear evidence for a band realignment that is accompanied by a transfer of electrons originally occupying the graphene's spin degenerate Dirac cones into $\alpha$-RuCl$_3$ band states with in-plane spin polarization. Left behind are holes in two separate Fermi pockets, only the dispersion of one of which is distorted near the Fermi energy due to spin selective hybridization with these spin polarized $\alpha$-RuCl$_3$ band states. This interpretation is supported by our DFT calculations. An unexpected damping of the quantum oscillations as well as a zero field resistance upturn close to the Néel temperature of $\alpha$-RuCl$_3$ suggests the onset of additional spin scattering due to spin fluctuations in the $\alpha$-RuCl$_3$.**

During the past decade, graphene has attracted immense interest due to its two-dimensional (2D) nature and intriguing Dirac band structure. It represents a highly tunable model system enabling the controlled study of a rich variety of phenomena that involve both the spin and valley degrees of freedom as relevant for spin- and valleytronics.[1,2] In addition, the encapsulation of graphene into van der Waals (vdW) heterostructures provides access to proximity effects to engineer the electronic structure, without compromising its structural integrity.[3–7] For instance, interfacial interaction between graphene and an adjacent material with a strong spin-orbit coupling (SOC) effect can enhance the intrinsically very weak SOC in graphene without inducing undesirable disorder, thus opening up a new venue for spin-based devices.[8–16] As another relevant proximity effect local spin generation and spin manipulation in graphene is achievable through an adjacent magnetic insulator via the magnetic exchange field.[17,18] Experimental and theoretical studies on graphene in proximity with magnetic insulators like EuO, EuS, or yttrium iron garnet (YIG)[19–22] have revealed novel correlated phenomena, such as the observation of a large magnetic exchange field and quantum anomalous Hall gaps.

Besides the above described proximity effects also sizable band hybridization can occur in such heterostructures. In particular, band hybridization signatures at higher binding energies without significant modification of graphene's bands close to the Fermi level have been detected by angle resolved photoemission spectroscopy (ARPES) on graphene/$MoS_2$ heterostructures.[23] Hybridization at binding energies between 3 and 6 eV causes the formation of several minigaps in the π bands of graphene.[23,24] While thus far ARPES measurements have been the primary tool to investigate such interfacial hybridization effects, the possibility to detect such effects also in electrical transport measurements when they appear near the Fermi energy would be highly desirable. The formation of new Fermi pockets as well as the deformation of existing ones can be captured by magneto-quantum oscillations of different periodicity reflecting contributions from multiple individual Fermi pockets each characterized by their own occupied density, effective mass and mobility. Scattering among electronic states belonging to different pockets as well as mutual electrostatic screening of the carriers in the different pockets may further enrich the transport behavior.[25] Hence, hybrid structures of graphene and a suitable magnetic insulator may introduce unprecedented transport phenomena.

In the present work, we explore modifications to graphene's electronic band structure induced by proximity of the layered 2D material α-RuCl$_3$. To this end, we resort to magnetotransport over a wide range of magnetic fields and temperatures. α-RuCl$_3$ is a 2D material which recently has attracted considerable attention owing to its close relation to spin liquid systems.[26,27,28] Previous studies have revealed that upon cooling below 10 K, the spins in α-RuCl$_3$ eventually get ordered in an in-plane zigzag pattern, corresponding to an antiferromagnetic phase.[29,30] As outlined in Figure 1a, the investigated graphene/α-RuCl$_3$ van der Waals heterostructures are composed of a bottom hBN layer, a graphene sheet, an α-RuCl$_3$ nanosheet of about 20 nm thickness, and finally an hBN cap layer on top. A particular challenge for the device fabrication (see Methods section) is that chemical damage of α-RuCl$_3$ occurs even for only short exposures to solvents such as acetone,[31] which are commonly used in e-beam lithography processes. To circumvent this issue, we have adjusted the sequence of fabrication steps. Specifically, e-beam lithography is performed first to create the Au contacts. An e-beam patterned resist serves as an etch mask for the contact pattern in the bottom multilayer hBN flake and subsequently the same resist is also used as lift-off mask to obtain self-aligned Au contacts fully integrated within the previously etched hBN layer. The device fabrication is completed by transferring a graphene/α-RuCl$_3$/hBN stack on top of this nearly planar surface with embedded contacts. A clean interface is achieved by 'brooming'[23] the bottom hBN layer with contact mode AFM prior to the transfer. In Figure 1b, an AFM image of a surface prepared in such manner is shown. Figure 1c displays an optical micrograph of a finished device. In the following, the electrical transport data recorded on one such device (referred to as device D1) are presented. Similar results were also obtained on a second device (D2), but their discussion is deferred to the Supporting Information.

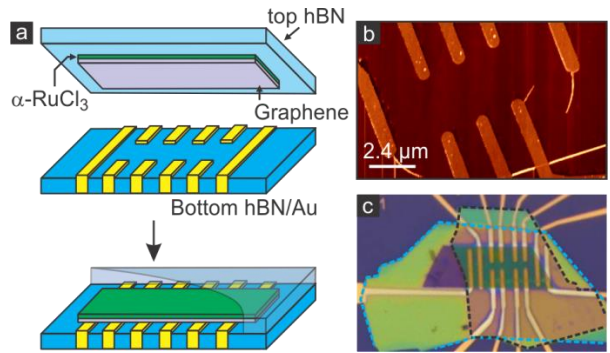

**Figure 1**. Device fabrication and layout. (a) Schematic illustration of the device fabrication process. The stack composed of a top-hBN layer, an α-RuCl$_3$ nanosheet and a graphene monolayer (top picture) is transferred onto gold electrodes, that are integrated within the bottom hBN (middle picture), in order to obtain the final graphene/α-RuCl$_3$ heterostructure device (bottom picture). Blue, green and purple color represents hBN, α-RuCl$_3$, and graphene, respectively. (b) AFM image of the bottom electrode structure after surface cleaning using contact mode AFM. (c) Optical image of a typical device with the polymer stamp on top of the stack being removed for clarity of the image. The dark dashed line demarcates the bottom hBN while the light blue dashed line encompasses the top hBN sheet.

First, the four-terminal resistance of such van der Waals heterostructures has been measured below 100 K where thin sheets of α-RuCl$_3$ that are not proximitized with graphene would become electrically insulating.[31] The dependence of the resistance on both temperature and back gate voltage is illustrated in Fig. 2. Panel a shows a color rendition across the explored parameter space spanned by temperature and gate voltage, whereas panel b and c depict single line traces at a set of selected fixed temperatures or gate voltages. There are two particularly striking differences in comparison to the behavior of pristine graphene. The resistance of only 4 to 8 $\Omega$ over the entire gate voltage and temperature range is unusually low. The observation of such anomalously low resistivity values for such heterostructures has recently been reported [32]. These very low values, together with the overall drop of the resistance towards more negative gate voltages, suggests dramatic p-type doping at the hybrid interface. Upon cooling below approximately 10 K the temperature dependence of the resistance deviates from the semi-metallic behavior of graphene, since a small upturn in resistance develops for all selected gate voltages (see Figure 2c). The upturn is more pronounced for positive gate voltages and the resistance traces above $V_g$ = +20 V fall almost on top of each other and can hardly be distinguished. One possible explanation for the resistance minimum observed in the resistance vs. temperature dependence (Figure 2c) relates to spin flip events in the $\alpha$-RuCl$_3$. These emerge already prior to the actual antiferromagnetic ordering at about 7 K, and hence may influence the electrical conductivity in this temperature range.[33] We note that both the dramatic drop in resistance as well as the upturn of the resistance at low temperature are generic and have been observed in all investigated devices. By contrast, the resistance feature in the gate voltage dependent traces of panel b around $V_g$ = +20 V did not appear in all devices (see for example additional data on device D2 in SI). Hence, it is not robust and may be disorder and sample specific.

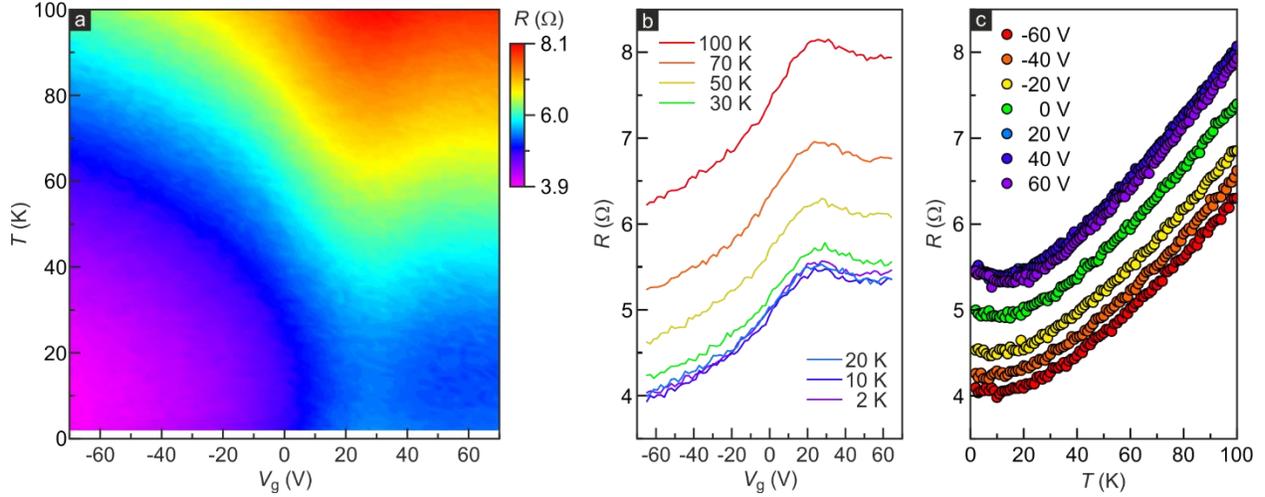

**Figure 2**. (a) Color map of the resistance of a graphene/α-RuCl$_3$ heterostructure (device D1) as a function of temperature and back gate voltage in the absence of an external magnetic field. (b) Resistance traces as a function of back gate voltage for several selected temperatures between 2 and 100 K. (c) Temperature evolution of the resistance at selected gate voltages between -60 and +60 V.

Magnetoresistance measurements are particularly instrumental to gain additional insight into the electronic structure at the graphene/α-RuCl$_3$ heterointerface. The longitudinal ($R_{xx}$) and Hall resistance ($R_{xy}$) as a function of applied *B*-field at $V_g = 10$ V and $T = 2$ K are plotted in Figure 3a and b, respectively. Quantum oscillations are apparent at *B*-fields as low as 4 T. The oscillating component of the longitudinal resistance, obtained by subtracting a smooth background ($\Delta R_{xx}$), is highlighted in Figure 3d by plotting the data as function of inverse *B*-field at 2 K and $V_g = 10$ V. The very fast period is indicative of a high carrier concentration as already concluded from the very low zero field resistance (Fig. 2). A beating pattern is also visible with a node for example near $B = 10$ T (Fig. 3a). This suggests contributions to the magnetoconductivity from multiple Fermi pockets of similar size. In general, each pocket hosting charge carriers with sufficiently high mobility will give rise to Shubnikov-de Haas (SdH) oscillations when its boundary in k-space at the Fermi energy describes a closed orbit. The characteristic $1/B$-periodicity yields in addition to a degeneracy factor the charge carrier density accommodated by the pocket. To evaluate how many different Fermi pockets are involved and how they evolve with the applied gate voltage, we have performed a FFT analysis of the oscillations. The resulting plot in Fig. 3e displays two well-distinguishable FFT peaks, reflecting the presence of two independent Fermi pockets. From the FFT frequency, the carrier density can be extracted by using $f_i = hn_i/g_i e$, where $h$ is the Planck constant, $e$ is the elementary charge, and $g_i$ is a possible degeneracy factor for pocket $i$. The obtained carrier concentrations normalized to the degeneracy factor, $n_i/g_i$, are on the order of $10^{13}$ cm$^{-2}$ (see Figure 3f). Their decrease with increasing gate voltage signifies $p$-type doping, in accordance with the gate-dependent resistance in Figure 2a and b. The difference in the carrier concentration between the two pockets is less than 10%.

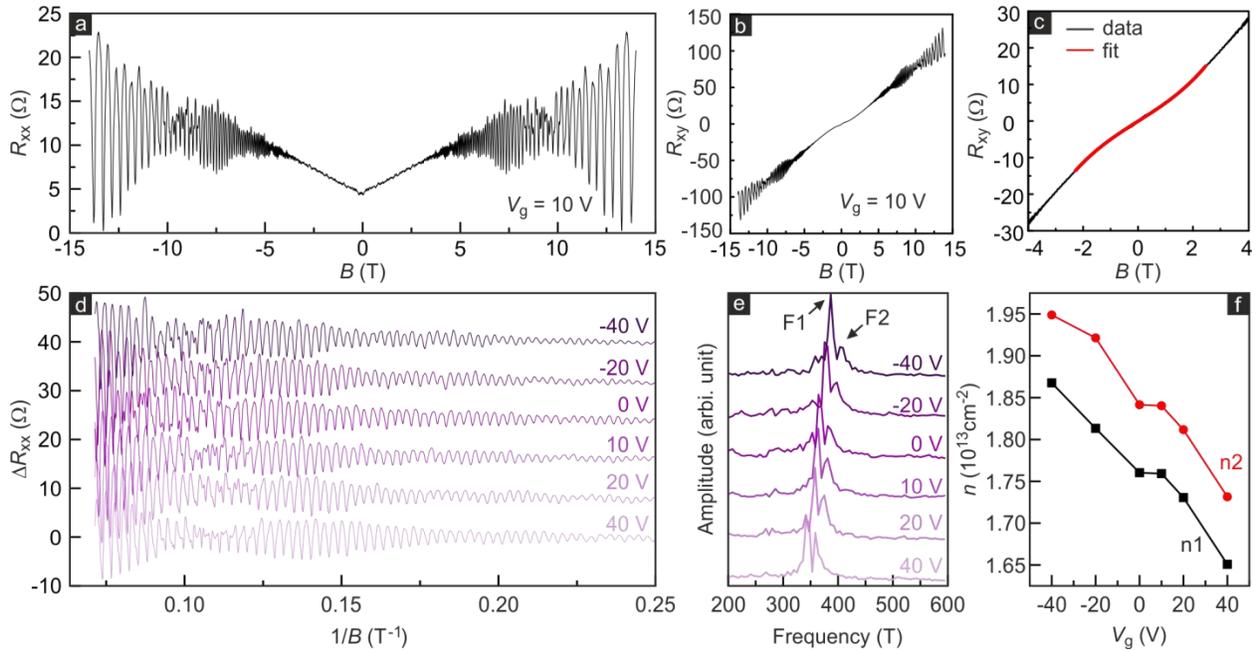

**Figure 3**. (a) Longitudinal resistance ($R_{xx}$) of the graphene/α-RuCl$_3$ heterostructure (device D1) as a function of *B*-field at $T = 2$ K and $V_g = 10$ V. (b,c) Hall resistance ($R_{xy}$) recorded at $V_g = 10$ V for (b) the full *B*-field range, and (c) zoom into the low magnetic field region. (d) Longitudinal resistance after subtracting a smooth background ($\Delta R_{xx}$), plotted as a function of inverse *B*-field for several gate voltages between -40 and 40 V. (e) FFT spectra of the SdH oscillations in panel (d). The same color code applies for panels (d) and (e). (f) Gate voltage-dependent carrier concentrations derived from peaks F1 and F2 in panel (e).

Finally, we turn our attention to the Hall curve in Figure 3b and c. Its slope confirms that holes are dominating transport. However, it also exhibits a clear non-linearity at fields below 2 T, i.e., the slope becomes less steep as zero field is approached. This is a Hall mark for the co-existence of both holes and electrons (see supplementary information S4) and suggests that apart from the two hole pockets producing quantum oscillations in the longitudinal resistance, also electrons are available which do not produce quantum oscillations. This may occur if they possess very poor mobility or if the associated Fermi contour does not form a closed orbit in k-space.

We resort to DFT calculations of the band structure at the graphene/α-RuCl$_3$-heterointerface in order to clarify how all these pieces of the puzzle fit together and what the origin of the magneto-transport features is. To properly describe the Mott insulating character of α-RuCl$_3$, we have employed the GGA+*U* scheme with implemented in-plane zigzag antiferromagnetic order.[34] The in-plane direction along which the zig-zag pattern forms (the inset of Fig. 4a), will be chosen as the *x*-direction. While further details of the calculations are relegated to the Method section, here we focus on their outcome. Figure 4a plots the electronic bands for the heterointerface. For ease of identification, states obviously associated with graphene's π-band Dirac cones have been colored green, while the flat bands in black originate from α-RuCl$_3$. The Fermi level corresponds to zero energy. An enlarged view of the band structure in the region marked by a circle is shown in panel b. States have been categorized as blue or red according to whether the projection of their spin along the *x*-direction is in the up state (red) or down state (blue). The size of the dots reflects the magnitude $S_x/[S_x^2+S_y^2+S_z^2]^{1/2}$ where $S_x$, $S_y$, and $S_z$ are the spin expectation values along the *x*, *y* and *z* direction. The work function difference[35,36,37] between graphene and α-RuCl$_3$ places an isolated electron-like band of α-RuCl$_3$ with a weak dispersion, spin up character and a minimum at the Γ–point just below the Fermi energy. This has important consequences, since the hybridization of this band with the π-band states of graphene splits the latter into states with either spin up or spin down character. The states with spin down character retain the original linear dispersion and are only affected by α-RuCl$_3$ bands at higher energies, followed by an anti-crossing with the α-RuCl$_3$ close to the Fermi level. This is seen more clearly in an even bigger enlargement of the band structure near the anti-crossing. The corresponding Fermi contours in panel d reveal the formation of a spin up and a spin down hole pocket centered around the K and K' points of the Brillouin zone. The spin up hole pocket is deformed and reduced in size as a result of the spin selective hybridization with the α-RuCl$_3$ band close to the Fermi energy. Concomitantly, an open Fermi surface on the right side of the anti-crossing in Fig. 4b containing the Γ-point appears and hosts electrons. The original C$_3$ symmetry inherited from α-RuCl$_3$'s atomic structure is reduced to C$_2$ symmetry due to the zigzag antiferromagnetic order of α-RuCl$_3$, and as a consequence the isotropic Fermi surface is transformed into an elongated pattern with an open Fermi contour colored red in Fig. 4d. It is noteworthy that the Dirac points are away from the α-RuCl$_3$ band and thus the band gap opening caused by the magnetic proximity effect is negligible (Fig. S7). This situation is different from what occurs at a graphene/CrI$_3$ heterointerface case where the states near the Dirac point of graphene directly hybridize with the unoccupied *d*-bands of CrI$_3$[38]. It is furthermore pertinent that appearance of two hole Fermi pockets, one of which with a distorted Fermi contour due to hybridization with a α-RuCl$_3$ band immediately close to the Fermi energy along with an electron pocket, as well as the corresponding dispersions are largely insensitive to the specific magnetic configuration of α-RuCl$_3$. This can be seen in Fig. S6 displaying band structure calculations assuming ferromagnetic ordering or a random spin configuration for α-RuCl$_3$ instead of antiferromagnetic order. In comparison, the magnetic configuration does have a strong impact on the spin character, i.e. both sign and magnitude of the spin projection onto the x-axis for the states belonging to each pocket. For the observed quantum oscillations and the Hall resistance the shape, area and dispersion of the pockets are relevant, whereas the spin projection remains hidden. In the remainder we will assume antiferromagnetic order as it has also been recently predicted for such a heterointerface.[37]

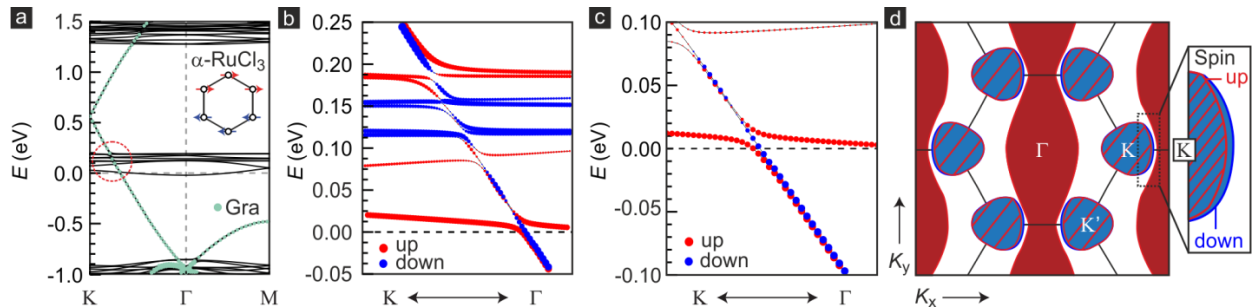

**Figure 4. Band structure of the graphene/α-RuCl$_3$ heterointerface obtained from DFT calculations**. (a) Overview of the bands between -1.0 and 1.5 eV. The states derived from graphene are represented by green dots. (b,c) Enlarged views on the band structure covering smaller energy windows between (b) -0.05 and 0.25 eV, and (c) -0.1 and 0.1 eV, respectively. Both plots reveal crossings and anti-crossing between graphene and α-RuCl$_3$ bands. The red and blue dots indicate whether states have a non-zero spin up or down projection of the spin along the $x$-direction defined as $S_x/[S_x^2+S_y^2+S_z^2]^{1/2}$ and the size of dot reflects the magnitude of this projection. (d) Map of the 2D Fermi surface contours. The red filled region represents the electron pocket with open Fermi contour. The blue filled hole pockets has spin down orientation, while the area hatched with red lines demarcates the hole pocket with spin up orientation. The hole pockets centered around the K and K' points originate from the modified graphene band, while the electron pocket at Γ stems from α-RuCl$_3$. The zoom on the right shows an enlarged view of the Fermi surface of the hole pockets at the K or K' points. The Fermi surface for states with spin down orientation is larger than that for the Fermi surface of up states, because the latter are affected by the hybridization with an α-RuCl$_3$ band with spin-up character.

This knowledge about the expected band structure modifications in α-RuCl$_3$-proximitized graphene puts us in a position to revisit the experimental data and identify unequivocally the origin of the quantum oscillations and the non-linearity in the Hall resistance. Specifically, to compensate for the work function difference, the Dirac point of graphene is shifted upward by more than 0.5 eV. The electrons normally occupying the Dirac cones populate a flat conduction band with spin up orientation that stems from the α-RuCl$_3$ and leave holes behind in two Fermi pockets with opposite in-plane spin orientation and closed contours centered around the K and K' points. They both possess a two-fold degeneracy ($g_1 = g_2 = 2$) that can be traced back to the valley degeneracy in graphene. Because the Fermi surface of the conduction band remains open, the electrons do not generate SdH oscillations, whereas the hole pockets produce two sets of SdH oscillations. The spin selective distortion of the hole spin up pocket due to a hybridization with the electron-like band accounts for the SdH-oscillations with a lower frequency than for the hole Fermi pocket that remains unaffected by this α-RuCl$_3$-band. The agreement between experimental observations and theoretical calculations is remarkable. Taking into account the degeneracy factor of 2, at a gate voltage of 10 V the larger pocket accommodates $n_2 = (1.84 \pm 0.012) \times 10^{13}$ cm$^{-2}$ holes, while the smaller one hosts $n_1 = (1.76 \pm 0.012) \times 10^{13}$ cm$^{-2}$. Using a two-carrier type model to fit the Hall resistance in Fig. 3c, we can extract the mobility and charge carrier densities of the holes and electrons that contribute to it (for details see Supporting Information). To this end, we fix the total hole density to the sum of the density in all hole pockets $n_h = n_1 + n_2$ (3.60 x 10$^{13}$ cm$^{-2}$) as determined from the SdH oscillations. A fit of the Hall curve recorded at $V_g = 10$ V then yields a hole mobility of 6000 cm$^2$/Vs. The non-linearity in the Hall resistance is best reproduced for a co-existing electron density equal to $n_e = (3.2 \pm 0.35) \times 10^{13}$ cm$^{-2}$ with a significantly lower mobility equal to 1900 cm$^2$/Vs (details of the fit procedure are provided in the Supporting Information). The open nature of the Fermi surface for these electrons accounts for the absence of SdH oscillations. To compare our experimental values with the calculated band structure, we have fixed the density of one hole pocket to the experimentally extracted value (by implementing an electric field of 50mV/Å perpendicular to the heterointerface in the calculation). Here $n_1$ has been fixed to 1.76 x 10$^{13}$ cm$^{-2}$. Subsequently, the second hole concentration, the electron concentration and Fermi level follow from the calculation of the band structure using DFT calculations. This yields another hole pocket with a carrier density of 1.67 x 10$^{13}$ cm$^{-2}$ and an electron pocket with a carrier density of 3.42 x 10$^{13}$ cm$^2$, in good agreement with the experimental values. One possibility that deserves consideration is whether the applied perpendicular magnetic field may induce a magnetic transition, resulting in a modified band hybridization. The AFM ordering in α-RuCl$_3$ is however in-plane and it has been shown in the literature that strong out of plane magnetic fields (above 15T), exceeding the field applied here, are needed to induce such a magnetic transition.[39]

Finally, we discuss the temperature dependence of the magnetoresistance oscillations of the graphene/α-RuCl$_3$ heterostructure. Exemplary traces up to 14 T are depicted in panel a of Fig. 5. The beating pattern is still visible up to 20 K, which is above the Néel temperature of α-RuCl$_3$. As mentioned above, the magnetic ordering of the α-RuCl$_3$ only affects the spin projection of the states (see supplementary information S7). However, the appearance of two hole pockets, one of which is deformed from its original Dirac cone shape near the Fermi energy due to hybridization with a band of α-RuCl$_3$, persists irrespective of the magnetic order. Only this is relevant for the emergence of the beating pattern as it results in a different occupation of the two pockets. The temperature evolution of the FFT peaks attributed to SdH oscillations stemming from hole pockets is plotted in Figure 5b. From the corresponding plot in Figure 5c, it can be seen that the oscillation amplitude does not monotonically increase upon cooling, but rather starts to slightly decrease below approximately 7-10 K. The same trend is also apparent in the raw data. The SdH oscillations at 4, 6, and 8 K are stronger than those at 2 K. The temperature onset of this unexpected damping of the SdH oscillation amplitude is close to the antiferromagnetic ordering (Néel) temperature of α-RuCl$_3$. This behavior hints toward extra spin scattering effects due to the onset of magnetic ordering in the α-RuCl$_3$ close to this temperature. In the same vein, such spin-related scattering could account for the resistance upturn at low temperatures (see Figure 2c).

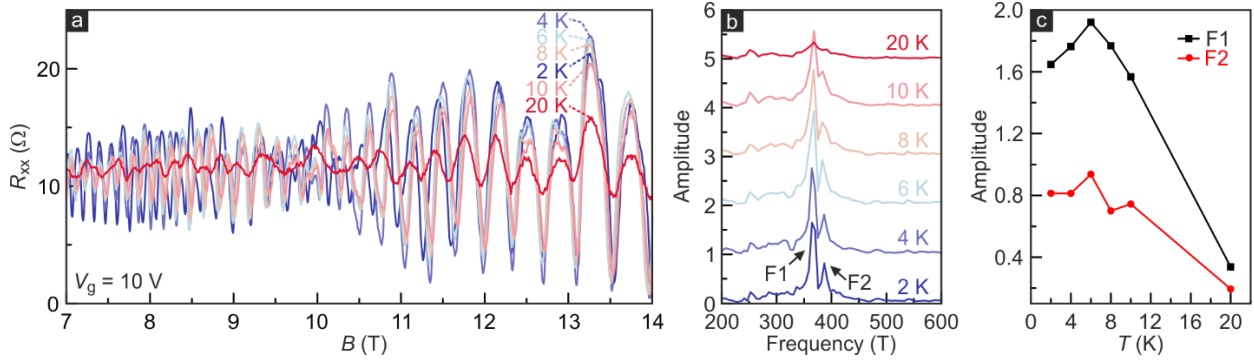

**Figure 5**. (a) $R_{xx}$ of the graphene/α-RuCl$_3$ heterostructure (device D1) as a function of B-field for several selected temperatures in the range between 2 and 20 K. (b) FFT spectra of the magnetoresistance in panel (a) at the same temperatures. (c) Temperature evolution of the FFT amplitude of the SdH oscillations, as extracted from the plot in panel (b).

In conclusion, the electrical and magnetotransport data that we gained on graphene/α-RuCl$_3$ heterostructure devices, in combination with our DFT calculations, provide convincing evidence for hybridization effects between the original graphene Dirac cones and the α-RuCl$_3$ bands. This hybridization occurs at different energies for the Dirac cone bands of opposite spin, resulting in two distinct hole pockets with different carrier concentrations near the Fermi level which causes a beating in the quantum oscillations. While this hybridization effect occurs irrespective of the specific magnetic ordering of α-RuCl$_3$, the presence or absence of magnetic order is reflected in the temperature dependent behavior of the magnetotransport as the Néel temperature of α-RuCl$_3$ is crossed.

**Methods:**

**Device Fabrication.** In a first step, hBN flakes were mechanically exfoliated onto Si substrates covered with a 300 nm thick thermally grown SiO$_2$ layer. A suitable hBN flake with a thickness of about 20-40 nm was selected when it exhibited a clean surface as judged from AFM topographic images. A resist mask patterned by e-beam lithography was then used to etch trenches into this hBN flake through a $SF_6$/Ar plasma (Oxford Plasma Pro100 Cobra system) with an $SF_6$ and Ar flow of 30 sccm and 25 sccm, respectively, a power of 12 W, and a chamber pressure of $5 \times 10^{-3}$ mbar. The same etch resist mask was then used to lift-off deposited electrodes. The latter were composed of 4 nm of Ti and a Au layer with a thickness matched to the thickness of the hBN flake in order to obtain a nearly planar surface and avoid mechanical stress when stacking the graphene/α-$RuCl_3$/hBN layers on top. These metals were thermally evaporated at a base pressure of $3 \times 10^{-8}$ mbar. The Au/hBN surface was then mechanically cleaned with the aid of contact mode AFM. Details of this cleaning procedure are provided elsewhere.[40] The van der Waals stack consisting of graphene/α-$RuCl_3$/top hBN was fabricated by the ELVACITE stamp method using motorized x-, y-, and z-stages. The stack was subsequently transferred on top of the Au/hBN area. The pick-up and release of the stack are controlled through adjustment of the substrate temperature.[41–43] Contrary to previously reported fabrication protocols using an ELVACITE stamp, the ELVCAITE layer is not removed at the final stage in order to prevent a chemical reaction between the acetone, required to remove this layer, and α-$RuCl_3$.

**Theoretical Electronic Band Structure Model for the Graphene/α-$RuCl_3$ Heterostructure**. The density functional theory calculations are carried out with the projected augmented plane-wave method[44,45] as implemented in the Vienna *ab initio* simulation package (VASP).[46] The Perdew-Burke-Ernzerhof functional of the generalized gradient approximation (GGA) is used for the description of exchange-correlation interactions among electrons[47], along with the van der Waals corrections that are needed to account for the interaction between graphene and α-RuCl$_3$.[48] We constructed a slab model with a 15 Å vacuum along the surface normal in which a (5×5) supercell of graphene is placed on a (√3×√3) supercell of α-RuCl$_3$. The energy cutoff for the plane-wave-basis expansion is set as 400 eV. We adopt the GGA +*U* scheme to incorporate the correlation effect on the *d*-shell of the Ru atoms.[34] To effectively reflect the substrate and thickness effects into our calculations, we also applied an external electric field of 50 mV/Å which pins one hole concentration and yields the Dirac point to be around 0.56 eV in energy as estimated in experiment.

**Associated Contents**

**Supporting Information**

Resistance measurements in the absence of an external magnetic field for device D2, magnetoresistance measurements for device D2, fast Fourier Transform analysis of SdH oscillations for device D2, temperature dependence of the SdH oscillations for device D2 and the robustness of the spin splitting (theory calculations).

**Author Information**

**Corresponding Author**

Soudabeh Mashhadi

Sh.Mashhadi@fkf.mpg.de

**Author Contributions**

S. M. and Y. K. contributed equally

**Notes**

The authors declare no competing financial interest.

**Acknowledgement**

J.H.S. acknowledges financial support from the graphene flagship. Y.K. is grateful for support from the Alexander Humboldt foundation. N. P. and J. K. were supported by the Basic Science Research Program through the NRF of Korea (NRF-2016R1D1A1B03931542). The growth of hexagonal boron nitride crystals was sponsored by the Elemental Strategy Initiative conducted by the MEXT, Japan and the CREST (JPMJCR15F3), JST. D.W. acknowledges the financial support by the German Science Foundation (DFG, WE6480/1).